\documentclass[aps,prd,twocolumn,showpacs,endfloats]{revtex4}
\usepackage{epsfig}
\usepackage{graphics}
\begin{document}

\def\vp{{\bf p}}
\def\ko{K^0}
\def\kb{\bar{K^0}}
\def\al{\alpha}
\def\ab{\bar{\alpha}}
\def\besub{\begin{subequations}}
\def\eesub{\end{subequations}}
\def\laball{\label{allequations}}
\def\be{\begin{equation}}
\def\ee{\end{equation}}
\def\bea{\begin{eqnarray}}
\def\eea{\end{eqnarray}}
\def\non{\nonumber}
\def\lab{\label}
\def\la{\langle}
\def\ra{\rangle}
\def\epp{\epsilon^{\prime}}
\def\vep{\varepsilon}
\def\to{\rightarrow}
\def\up{\uparrow}
\def\dw{\downarrow}
\def\ms{\overline{\rm MS}}
\def\ums{{\mu}_{_{\overline{\rm MS}}}}
\def\u{\mu_{\rm fact}}

\def\pr{{Phys. Rev.}~}
\def\ijmp{{Int. J. Mod. Phys.}~}
\def\jp{{J. Phys.}~}
\def\mpl{{Mod. Phys. Lett.}~}
\def\prp{{Phys. Rep.}~}
\def\prl{{Phys. Rev. Lett.}~}
\def\pl{{Phys. Lett.}~}
\def\np{{Nucl. Phys.}~}
\def\ppnp{{Prog. Part. Nucl. Phys.}~}
\def\zp{{Z. Phys.}~}
\def\epj{{Eur. Phys. J.}~}

\title{Intrinsic Charm Flavor and Helicity Content in the Proton$^*$\\
}
\author{Xiaotong Song}
\email{xs3e@virginia.edu}
\thanks{\\
$^*$ Invited paper to the 3rd Circum-Pan-Pacific Symposium on 
``High Energy Spin Physics'', Beijing, October 8$-$13, 2001.}
\thanks{\\
$^{**}$ also Department of Physics, Zhejiang University, 
Hangzhou, Zhejiang, P. R. China}

\affiliation{Institute of Nuclear and Particle Physics\\
Jesse W. Beams Laboratory of Physics\\
Department of Physics, University of Virginia,
Charlottesville, VA 22904, USA\\}

\begin{abstract}
Contributions to the quark flavor and spin observables from the 
intrinsic charm in the proton are discussed in the SU(4) quark 
meson fluctuation model. Our results suggest that the probability 
of finding the intrinsic charm in the proton is less than $1\%$. 
The intrinsic charm helicity is small and negative, $\Delta c
\simeq -(0.003\sim 0.015)$. The fraction of the total quark 
helicity carried by the intrinsic charm is less than $2\%$, and
$c_\up/c_\dw=35/67$. 
\end{abstract}
\par
\maketitle

\leftline{\bf 1. Introduction}

The {\it intrinsic} heavy quark component in the nucleon wave function 
has been suggested by many authors long time ago \cite{stan81,dg77}. 
This component, created from the quantum fluctuations associated with 
the bound state hadron dynamics, exists in the hadron over a long time 
independent of any external probe momentum. The probability of finding 
the intrinsic heavy quarks in the hadron is completely determined by 
nonperturbative mechanisms. On the other hand, the {\it extrinsic} heavy 
quarks are created on a short time scale in association with a large 
transverse momentum reaction and their distributions can be derived 
from QCD bremsstrahlung and pair production processes, which lead to 
standard QCD evolution. At the scale $m_c^2$ or lower, we only need to 
consider the intrinsic charm (IC) contribution. An interesting question 
is what will be the size of the IC contribution to the flavor and spin 
observables of the proton if the IC does exist. Since there is no direct 
experimental data of the IC content, one has to resort to the nucleon 
models (see e.g. \cite{dg77,3}) or combination of using the model and 
analysing the DIS data to obtain some information of the IC contribution 
(see e.g. \cite{4}).

Although the SU(3) chiral quark model with symmetry breaking provides 
a useful nonperturbative tool in describing the quark spin, flavor 
\cite{song9705} and orbital structure \cite{song0012}, the model is 
quite {\it unnatural} from the point of view of the standard model. 
According to the symmetric GIM model \cite{gim70}, one should deal with 
the weak axial current in the framework of SU(4) symmetry. It implies 
that the charm quark should be included in determining the spin, flavor 
and orbital structure of the nucleon. In an earlier report 
\cite{songictp98}, the author has suggested an extended SU(4) version 
of the chiral quark model and presented some preliminary results. In 
the chiral quark model or more precisely the quark meson fluctuation 
model (some earlier works on this model see e.g. \cite{9}), the nucleon 
structure is determined by its valence quark configuration and all 
possible quantum fluctuations of valence quarks into quarks plus mesons. 
In the SU(4) model, the charm or anti-charm quarks reside in the charmed 
mesons which are created by nonperturbative quantum quark-meson 
fluctuations. Hence these charm or anticharm quarks are essentially 
$intrinsic$. 
\bigskip

\leftline{\bf 2. SU(4) model with symmetry breaking}

In the framework of SU(4) quark model, there are sixteen pseudoscalar 
mesons, a 15-plet and a singlet. In this paper, the contribution of
the SU(4) singlet will be neglected. The effective Lagrangian describing 
interaction between quarks and the mesons is
\be
{\it L}_I=g_{15}{\bar q}\pmatrix{{G}_u^0
& {\pi}^+ & {\sqrt{\epsilon}}K^+ & {\sqrt{\epsilon_c}}{\bar D}^0 \cr 
{\pi}^-& {G}_d^0& {\sqrt{\epsilon}}K^0 &{\sqrt{\epsilon_c}}D^-\cr
{\sqrt{\epsilon}}K^-& {\sqrt{\epsilon}}{\bar K}^0&{G}_s^0 &
{\sqrt{\epsilon_c}}{D}_s^-\cr 
{\sqrt{\epsilon_c}}{D}^0 & {\sqrt{\epsilon_c}}{D}^+
& {\sqrt{\epsilon_c}}{D}_s^+ &{G}_c^0 \cr }q, 
\ee
where 
$D^+=(c\bar d)$, $D^-=(\bar cd)$, $D^0=(c\bar u)$, $\bar D^0=(\bar cu)$,
$D_s^+=(c\bar s)$, and $D_s^-=(\bar cs)$. The neutral charge components
${G}_{u(d)}^0$ and ${G}_{s,c}^0$ are defined as
\be
{G}_{u(d)}^0=+(-){{\pi^0}\over{\sqrt 2}}+
{\sqrt{\epsilon_{\eta}}}{{\eta^0}\over{\sqrt 6}}+
{\zeta'}{{\eta'^0}\over{4\sqrt 3}}-{\sqrt{\epsilon_c}}{{\eta_c^0}\over 4}
\ee
\be
{G}_s^0=-{\sqrt{\epsilon_{\eta}}}{{2\eta^0}\over{\sqrt 6}}+
{\zeta'}{{\eta'^0}\over{4\sqrt 3}}-{\sqrt{\epsilon_c}}{{\eta_c^0}\over 4}
\ee
\be
{G}_c^0=-{\zeta'}{{3\eta'^0}\over {4\sqrt 3}}+
{\sqrt{\epsilon_c}}{{3\eta_c^0}\over 4}
\ee
with $\pi^0=(u\bar u-d\bar d)/\sqrt 2$, $\eta^0=(u\bar u+d\bar d-2s\bar s)
/\sqrt 6$, $\eta'^0=(u\bar u+d\bar d+s\bar s)/\sqrt 3$, and 
$\eta_c^0=(c\bar c)$. Similar to the SU(3) case, we define $a\equiv 
|g_{15}|^2$, which denotes the transition probability of splitting 
$u(d)\to d(u)+\pi^{+(-)}$. Hence $\epsilon a$, $\epsilon_{\eta} a$ 
and $\epsilon_c a$ denote the probabilities of splittings 
$u(d)\to s+K^{-(0)}$, $u(d)\to u(d)+\eta^{(0)}$ and 
$u(d)\to c+{\bar D}^{0}(D^-)$ respectively. If the breaking effects are 
dominated by the mass differences, we expect 
$0~<~\epsilon_{c}~<~\epsilon~\simeq~\epsilon_{\eta}~<~1$. 

In addition to the allowed fluctuations discussed in the SU(3) case, 
a valence quark ($u$ or $d$ in the proton) is now allowed to split up 
or fluctuate to a recoil charm quark and a charmed meson. For example, 
a valence u-quark with spin-up, the allowed fluctuations are
\bea
u_{\up}&\to& d_{\dw}+\pi^+,~~
u_{\up}\to s_{\dw}+K^+,~~
u_{\up}\to u_{\dw}+{G}_u^0,\\
u_{\up}&\to& c_{\dw}+{\bar D}^0,\\
u_{\up}&\to& u_{\up}.
\eea
Similarly, one can list the allowed fluctuations for $u_{\dw}$,
$d_{\up}$, $d_{\dw}$, $s_{\up}$, and $s_{\dw}$. 
Similar to the SU(3) \cite{song0012} case, the spin-up and spin-down 
quark or antiquark contents in the proton, up to first order of the 
quantum fluctuation, can be calculated.
\bigskip

\leftline{\bf 3. Quark flavor and spin contents}

We note that the quark {\it flips} its spin in the splitting processes 
$q_{\up,(\dw)}\to q_{\dw,(\up)}$+meson, i.e. processes in (5) and (6),
but not in $u_{\up}\to u_{\up}$. The quark helicity non-flip contributions
in the splitting processes (5) and (6) are neglected, which is the 
basic assumption in the model.

\leftline{\quad \bf 3.a. Flavor content in the nucleon}

The quark and antiquark flavor contents are 
\be
u=2+\bar u,~~~d=1+\bar d,~~~s=0+\bar s,~~~c=0+\bar c,~~~
\ee
\be
\bar u=a[1+\tilde A^2+2(1-\tilde A)^2],~~~
\bar d=a[2(1+\tilde A^2)+(1-\tilde A)^2],~~~
\ee
\be
\bar s=3a[\epsilon+\tilde B^2],~~~\bar c=3a[\epsilon_c+\tilde D^2],
\ee
where $\tilde A$, $\tilde B$, $\tilde C$, and $\tilde D$ are defined 
similar to those in the SU(3) case. From (9), one obtains
\be
\bar u/\bar d=1-6\tilde A/[(3\tilde A-1)^2+8],
\ee
and
\be
\bar d-\bar u=2a\tilde A.
\ee
Similarly, one can obtain $2\bar c/(\bar u+\bar d)$, $2\bar c/(u+d)$, 
$2\bar c/\sum(q+\bar q)$ and other flavor observables.
One remark should be made here. Defining the ratio, $r\equiv\bar u/\bar 
d$, we obtain, from Eqs. (11) and (12), 
\be
1/2~\leq ~\bar u/\bar d~ \leq~ 5/4, 
\ee
which seems to be consistent with the experimental data shown in Table I.

\leftline{\quad \bf 3.b. Helicity content in the nucleon}

Similarly, we obtain
\be
\Delta u=(4/3)[1-a(\epsilon+\epsilon_c+2f)]-a, 
~~~~\Delta c=-a\epsilon_c,
\ee
\be
\Delta d=(-1/3)[1-a(\epsilon+\epsilon_c+2f)]-a,
~~~~\Delta s=-a\epsilon_s,
\ee
(where $f$ is generalization of $f_{SU(3)}$ defined in \cite{song9705}) 
and 
\be
\Delta{\bar q}=0,~~~~~~({\bar q}={\bar u},{\bar d},{\bar s},{\bar c}). 
\ee
Several remarks are in order.
\begin{itemize}
\item{In the splitting process $u_{\up(\dw)}~\to~c_{\dw(\up)}+\bar D^0$, 
the anticharm resides only in the charmed meson, e.g. $\bar D^0(\bar 
c,u)$. The probabilities of finding $\bar c_{\up}$ and $\bar c_{\dw}$ 
are equal in the spinless charmed meson. Therefore $\Delta\bar c=0$. 
Similar discussion in the SU(3) case has led to $\Delta\bar q=0$ for 
$\bar q=\bar u,\bar d,\bar s$. The DIS data \cite{abe97} seems to 
support this prediction.}
\item{The charm quark helicity $\Delta c$ is {\it nonzero} as far as
$\epsilon_c$ is nonzero. Analogous to the strange quark helicity,
$\Delta c$ is definitely {\it negative}, because in the splitting
processes, $u_{\up(\dw)}~\to~c_{\dw(\up)}+\bar D^0$ and
$d_{\up(\dw)}~\to~c_{\dw(\up)}+D^-$, more $c_{\dw}$ is created than 
$c_{\up}$, because of the probability of finding the valence $u$-quark
in the zeroth approximation, $n^{(0)}_p(u_{\up})$, is dominant.}
\item{From (10) and (14), using ${\tilde D}^2={\epsilon_c}/16$,
one can see that the ratio 
\be
\Delta c/\bar c=-16/51
\ee
is a constant, which {\it does not depend on} any splitting parameters. 
This is a special prediction for the charm flavor in the SU(4) quark 
meson model. Combining (17) and (16), one obtains $c_\up/c_\dw=35/67$.
For the strangeness, if $\zeta'=0$, one has similar result, i.e.
$\Delta s/\bar s=-3/10$ is also a constant and $s_\up/s_\dw=7/13$.} 
\end{itemize}
\bigskip

\leftline{\bf 4. Numerical results and discussion.}

Since the effect arising from splitting (6) is smaller than those 
from (5), we expect the values of parameters $a$ and $\epsilon$ 
in SU(4) should be very close to those used in SU(3) version, where 
$a=0.145$, $\epsilon=0.46$. We choose $a=0.143$, $\epsilon=0.454$, 
and leave $\epsilon_c$ as a $variable$, then the quark flavor and 
helicity contents can be expressed as functions of $\epsilon_c$. 
To determine the value of $\epsilon_c$, we use the low energy hyperon 
$\beta$-decay data \cite{pdg00}, $\Delta_3=1.2670\pm 0.0035$. We find  
\be
\epsilon_c\simeq 0.06\pm 0.04.
\ee
Using only $three$ parameters, $\{ a, \epsilon, \epsilon_c \}$,
the flavor and spin observables are calculated and listed in Table I 
and Table II respectively. For comparison, we also list the existing 
data and results given by SU(3) description and other models or analyses. 
One can see that the model satisfactorily describes almost all the 
existing data and also gives some new predictions. Several remarks are 
in order:
\begin{itemize}
\item{The theoretical uncertainties shown in the quantities 
in Tables I and II arise from the uncertainty of $\epsilon_c$ in (11).
If the observable does not depend on $\epsilon_c$, such as $\bar d-
\bar u$, $\bar d/\bar u$, $2\bar s/(\bar u+\bar d)$, etc., there is no 
uncertainty for them. Two special quantities $\Delta c/\bar c$ and 
$\Delta s/\bar s$ are also independent of $\epsilon_c$ (see
Table II).}
\item{The SU(4) version predicts the IC component in the proton, 
${2\bar c}/\sum(q+\bar q)\simeq 1\%$, which agrees with the predictions 
given in \cite{dg77} and [4e], and is also close to the those given in 
[3a], [3d] and [4f]. We note that the IC component is almost one order 
of magnitude smaller than the intrinsic strange component ${2\bar 
s}/\sum(q+\bar q)$. }
\item{Using similar approach given in a previous work (see Eq. (3.6) in 
[9c], we can show that 
$<2x\bar c(x)>/<\sum[xq(x)+x\bar q(x)]>$ is smaller than $2\bar 
c/\sum(q+\bar q)$, where $q(\bar q)\equiv\int_0^1dxq(\bar q)(x)$, and 
$<xq(\bar q)(x)>\equiv\int_0^1dxxq(\bar q)(x)$. It implies that
the fraction of the total quark momentum carried by the charm and 
anticharm quarks is less than $1\%$.} 
\item{The prediction of intrinsic charm polarization, $\Delta c\simeq
-0.009\pm 0.006$ is close to the result $\Delta c=-0.020\pm 
0.005$ given in the instanton QCD vacuum model [3c]. Our result is 
smaller than that given in [3b] ($\Delta c\simeq -0.3$). However, the 
size of $\Delta c\simeq -5\cdot 10^{-4}$ given in [3d] is even smaller. 
Hence further investigation in this quantity is needed.} 
\item{Taking $\epsilon_c\simeq 0.06$, one has
$\Delta c/\Delta\Sigma\simeq -0.02$. This is consistent with the 
prediction given in [3c], but smaller than that given in [3b].
Combining with the fractions of the light quark helicities, we have 
$\Delta u/\Delta\Sigma\simeq 2.17$, $\Delta d/\Delta\Sigma\simeq -0.99$, 
$\Delta s/\Delta\Sigma\simeq -0.16$, and $\Delta c/\Delta\Sigma\simeq 
-0.02$. One can see that the $u$-quark helicity is $positive$ (parallel 
to the nucleon spin) and about two times larger than the total quark 
helicity $\Delta\Sigma$. However, the $d$-, $s$-, and $c$-helicities 
are all $negative$ (antiparallel to the nucleon spin), and their sizes 
are decreased as 
\be
\Delta d~:~\Delta s~:~\Delta c~\simeq~1~:~10^{-1}~:~10^{-2}.
\ee
Compare to the strange helicity $\Delta s$, the IC helicity is one order 
of magnitude smaller.}
\end{itemize}
 
To summarize, we have discussed the IC contribution in the SU(4) quark 
meson model with symmetry breaking. Our results suggest that the 
probability of finding the IC in the proton is in the range $0.003\sim
0.019$, and the IC helicity is small and negative, $\Delta c\simeq 
-(0.003\sim 0.015)$. The fraction of the total quark helicity carried 
by the intrinsic charm is also small, $\Delta c/\Delta\Sigma\simeq 
-(0.007\sim 0.035)$. 
\bigskip

\acknowledgments

This work was supported in part by the U.S. DOE Grant, the Institute 
of Nuclear and Particle Physics, Department of Physics, University of 
Virginia, and the Commonwealth of Virginia.

\widetext
\begin{table}
\nopagebreak
\caption{Quark Flavor Observables.}
\begin{tabular}{|c|c|c|c|} 
\hline
Quantity & Data   & SU(4) [This paper]
& SU(3) \cite{song0012}\\
\hline 
$\bar d-\bar u$ & $0.110\pm 0.018$\cite{peng98} & 0.111 &  0.143  \\
                & $0.147\pm 0.039$\cite{am94} &       &    \\
\hline 
${{\bar u}/{\bar d}}$ &$[\bar u(x)/\bar d(x)]_{0.1<x<0.2}
=0.67\pm 0.06$\cite{peng98} & 0.71 &  0.64\\
&$[\bar u(x)/\bar d(x)]_{x=0.18}=0.51\pm 0.06$\cite{na51} & 
& \\ 
\hline
${{2\bar s}/{(\bar u+\bar d)}}$ & $<2x\bar s(x)>/<x(\bar
u(x)+\bar d(x))>=0.477\pm 0.051$\cite{ba95}& 0.66& 0.76\\
\hline
${{2\bar c}/{(\bar u+\bar d)}}$ & $-$ & $0.083\pm 0.055$ &0 \\
\hline
${{2\bar s}/{(u+d)}}$ & $<2x\bar s(x)>/
<x(u(x)+d(x))>=0.099\pm 0.009$\cite{ba95}&0.118 &0.136\\
\hline
${{2\bar c}/{(u+d)}}$ & $-$ & 0.015$\pm 0.010$ &0\\
\hline
$(s+\bar s)/\sum(q+\bar q)$ & 
$<2x\bar s(x)>/\sum<x(q(x)+\bar q(x))>=0.076\pm 0.022$\cite{ba95} 
& 0.090$\pm 0.001$ &0.103\\
       &$0.10\pm 0.06$\cite{gls91}  &     & \\
       &$0.15\pm 0.03$\cite{dll96}  &     & \\
\hline
$(c+\bar c)/\sum(q+\bar q)$ & 0.03~[4f]$^*$&0.011$\pm 0.008$ & 0\\
& 0.02~\cite{dg77}$^*$ &  &\\
& 0.01~[4e]$^*$ &  & \\
& 0.005~[3a,3d]$^*$&  & \\
\hline
${{\sum\bar q}/{\sum q}}$ & $\sum<x\bar q(x)>/\sum<xq(x)>=0.245\pm 
0.005$\cite{ba95}&$0.230\pm 0.004$
&0.231\\
\hline
\end{tabular}
\end{table}

\widetext
\begin{table}
\nopagebreak
\caption{Quark Spin Observables}
\begin{tabular}{|c|c|c|c|}\hline
\hline
Quantity & Data   &  SU(4) [This paper]&SU(3) \cite{song0012}\\
\hline 
$\Delta u$ & $0.85\pm 0.04$\cite{adams97} & $0.871\pm 0.009$ &0.863\\
$\Delta d$&$-0.41\pm$0.04\cite{adams97} &$-0.397\pm 0.002$ &$-0.397$\\
$\Delta s$&$-0.07\pm$0.04\cite{adams97}&$-0.065\pm 0.000$ &$-$0.067\\
\hline
$\Delta c$ & $-0.020\pm 0.004$~[3c]$^*$&$-0.009\pm 0.006$ &0\\
& $-0.3$~[3b]$^*$ &  & \\
& $-5\cdot 10^{-4}$~[3d]$^*$ &  & \\
\hline
$\Delta\Sigma$/2 & $0.19\pm 0.06$\cite{adams97}& $0.200\pm 0.006$ &0.200\\
\hline
$\Delta\bar u$, $\Delta\bar d$ & $-0.02\pm 0.11$\cite{adeva96}&0 &0\\
\hline
$\Delta\bar s$, $\Delta\bar c$ & $-$&0 &0\\
\hline
$\Delta c/\Delta\Sigma$ & 
$-0.08\pm 0.01$~[3b]$^*$ & $-0.021\pm 0.014$ &0 \\
  & $-0.033$ ~[3c]$^*$& &  \\
\hline
$\Delta s/{\bar s}$ & $-$ & $-3/10$ & $-0.269$  \\
\hline
$\Delta c/{\bar c}$ & $-$ & $-16/51$ & $-$  \\
\hline 
$c_\up/c_\dw$ & $-$ & $35/67$ & $-$  \\
\hline 
$s_\up/s_\dw$ & $-$ & $7/13$ & $\simeq 0.58$  \\
\hline 
$\Gamma_1^p$ & $0.136\pm 0.016$\cite{adams97} & $0.143\pm 0.002$ &0.142\\
$\Gamma_1^n$ & $-0.041\pm 0.007$\cite{abe97} &$-0.042\pm 0.001$ 
&$-0.042$\\
\hline 
$\Delta_3$&1.2670$\pm$0.0035\cite{pdg00}&$1.268\pm 0.010$ &1.260\\
$\Delta_8$& 0.579$\pm$ 0.025\cite{pdg00}&$0.605\pm 0.006$ &0.600 \\
\hline
\end{tabular}
\end{table}


\begin{thebibliography}{99}

\bibitem{stan81}
	S.~J.~Brodsky, P.~Hoyer, C.~Peterson and N.~Sakai, {\pl} {\bf
B93}, 451 (1980);~ S.~J.~Brodsky, and C.~Peterson, {\pr} {\bf D23}, 2745 
(1980).

\bibitem{dg77} 
	J.~F.~Donoghue and E.~Golowich, {\pr} {\bf D15}, 3421 (1977);
H.~He, X.~Zhang, and Y. Zhao, {\it High Energy Phys. Nucl. Phys.} {\bf 7},
626 (1983).

\bibitem{3}
	a) T.~Hatsuda, and T.~Kunihiro, {\prp} {\bf 247}, 221  
(1994); b) A.~Blotz and E.~Shuryak, {\pl} {\bf B439}, 415 (1998);
c) F.~Araki, M.~Musakhanov and H.~Toki, hep-ph/9808290;
d) M.~V.~Polyakov, A.~Schafer, and O.~V.~Teryaev, {\pr} {\bf D60}, 
051502 (1999);~ M.~Franz, M.~V.~Polyakov, and K.~Goeke, {\pr} {\bf D62},
074024 (2000).

\bibitem{4}
	a) E.~Hoffmann and R.~Moore, {\zp} {\bf C20}, 71 (1983);
b) B.~W.~Harris, J.~Smith and R.~Vogt, {\np} {\bf B461}, 181 (1996);
c) R.~Vogt and S.~J.~Brodsky, {\np} {\bf B478}, 311 (1996);
d) G.~Ingelman and M.~Thunman, {\zp} {\bf C73}, 505 (1997);
e) Y.~A.~Golubkov, {\it Phys. Atom. Nucl.} {\bf 63}, 606 (2000);
f) S.~J.~Brodsky, and I.~Schmidt, {\pr} {\bf D43}, 179 (1991);
g) J.~Blumlein and W.~L.~van Neerven, {\pl} {\bf B450}, 417 (1999).

\bibitem{song9705} 
	X.~Song, {\pr} $\bf D57$, 4114 (1998).

\bibitem{song0012} 
	X.~Song, {\ijmp} {\bf A16}, 2673 (2001).

\bibitem{gim70} 
	S.~L.~Glashow {\it et al.}, {\pr} {\bf D2}, 2185 (1970).

\bibitem{songictp98} 
	X.~Song, Internal Report, ICTP, June (1998).

\bibitem{9} 
     	a) E.~J.~Eichten, I.~Hinchliffe and C.~Quigg, {\pr} {\bf D45}, 
{2269} (1992); {\pr} {\bf D47}, 747 (1993);
b) T.~P.~Cheng and L.-F.~Li, {\prl} {\bf 74}, {2872} (1995);
c) X.~Song, J.~S.~McCarthy and H.~J.~Weber, {\pr} {\bf D55}, 2624 (1997).

\bibitem{abe97}
	K.~Abe, {\it et al.}, {\prl} {\bf 79}, 26 (1997).

\bibitem{pdg00}
	Particle Data Group, D.~E.~Groom, {\it et al.}, {\epj} 
{\bf C15}, 1 (2000).

\bibitem{peng98}
	J.~C.~Peng, {\it et al.}, E866/NuSea Collaboration, {\pr} {\bf 
D58}, 092004 (1998).

\bibitem{am94} 
	 P.~Amaudruz {\it et al.}, {\prl} {\bf 66}, {2712} (1991);
M.~Arneodo {\it et al.}, {\pr} {\bf D50}, R1, (1994).

\bibitem{na51}
	A.~Baldit, {\it et al.}, NA51 Collaboration, {\pl} {\bf B332},
244 (1994).

\bibitem{ba95}
	A.~O.~Bazarko, {\it et al.}, {\zp} {\bf C65}, 189 (1995).

\bibitem{gls91}
	J.~Gasser, H.~Leutwyler and M.~E.~Saino, {\pl} {\bf B253},
252 (1991).

\bibitem{dll96}
	S.~J.~Dong, J.~F.~Lagae and K.~F.~Liu, {\pr} {\bf D54},
5496 (1996).

\bibitem{adams97}
	P.~Adams, {\it et al.}, {\pr} {\bf D56}, 5330 (1997).

\bibitem{adeva96}
	B.~Adeva, {\it et al.}, {\pl} {\bf B369}, 93 (1996).

\end{thebibliography}
\end{document}